\documentclass{ptephy_v1}

\preprintnumber{XXXX-XXXX} 

 \usepackage{graphicx}
\usepackage{amsmath}
\usepackage{lipsum}
\usepackage{amsmath,bm}
\def\vect#1{\mbox{\boldmath $#1$}}

\newcommand{\oo}{${{^{16}}{\rm O}}$}

\newcommand{\nene}{${^{20}{\rm Ne}}$}
\newcommand*{\mycdot}{\kern-.2em\cdot\kern-.2em}

\begin{document}
\title{New trial wave function for nuclear cluster structure of nuclei}
 

\author{Bo Zhou}
\affil{Institute for International Collaboration, Hokkaido University, Sapporo 060-0815, Japan}
\affil{Department of Physics, Hokkaido University, Sapporo 060-0810, Japan \email{bo@nucl.sci.hokudai.ac.jp}}


\begin{abstract}%
A new trial wave function is  proposed for nuclear cluster physics, in which an exact solution to the long-standing center-of-mass problem is given. In the new approach, the widths of the single-nucleon Gaussian wave packets and the widths of the relative Gaussian wave functions describing correlations of nucleons or clusters are treated as variables in the explicit intrinsic wave function of the nuclear system. As an example, this new wave function was applied to study the typical ${^{20}{\rm Ne}}$ ($\alpha$+${{^{16}}{\rm O}}$) cluster system. By removing exactly the spurious center-of-mass effect in a very simple way, the energy curve of ${^{20}{\rm Ne}}$ was obtained by the variational calculations with the width of $\alpha$ cluster, the width of ${{^{16}}{\rm O}}$ cluster, and the size parameter of the nucleus. They  are considered as the three crucial variational variables in describing the  ${^{20}{\rm Ne}}$ ($\alpha$+${{^{16}}{\rm O}}$) cluster system. This shows that the new wave function can be a very interesting new tool for studying many-body and cluster effects in nuclear physics.
\end{abstract}

\subjectindex{D05, D10, D11}

\maketitle

The many-body problem in quantum physics is one of the great challenges in various fields, e.g., quantum chemistry~\cite{shavitt_many-body_2009}, condensed matter physics~\cite{nozieres_theory_1997}, and nuclear physics ~\cite{ring_nuclear_2004}, in which the non-negligible correlations are encoded in the extremely complexity of the many-body wave function. In many-body theory~\cite{kamada_benchmark_2001,bloch_many-body_2008,carleo_solving_2017},  a good trial wave function is a key ingredient for describing a physical system.
However, the construction of a suitable many-body trial wave function has never been easy. The accumulating evidences strongly suggest that a good many-body trial wave function should not only have  clear physical significance but also should possess a quite suitable form for practical massive calculations.

The nuclear system is a natural laboratory for the quantum many-body problem. In recent years, a variety of true quantum many-body wave functions were proposed for the study of nuclear dynamics, most of which are represented as Slater determinants but employ different degrees of freedom for the dynamical evolution of nuclear systems~\cite{friman_cbm_2011}. Antisymmetrized Molecular Dynamics (AMD)~\cite{kanada-enyo_clustering_1995,itagaki_new_2003,kimura_antisymmetrized_2016} is one representative of many-nucleon wave functions, in which the spatial part of the single-particle basis is expressed by the Gaussian packet  $\exp[-(\bm{r}_i-\bm{Z}_i)^2/(2b^2)] $ with common harmonic oscillator width $b$ and different position variables \{$\bm{Z}_i$\}. 
By superposing many Slater determinants with the obtained optimum values \{$\bm{Z}_i$\}, the correlations of nucleons can be obtained and the clustering and mean-field effects in light nuclei are studied well in the framework of AMD~\cite{kanada-enyo_antisymmetrized_2012}.
Fermionic Molecular Dynamics (FMD)~\cite{feldmeier_fermionic_1990,neff_cluster_2004} is another more sophisticated many-body wave function. In FMD, not only the Gaussian wave packet centers  \{$\bm{Z}_i$\} but also the widths \{$b_i$\} become dynamical variables rather than the constant parameter.
The importance of dynamics wave packet widths has been discussed in many works~\cite{sunkel_generator_1976,colonna_spinodal_1998,maruyama_extension_1996,tang_resonating-group_1978}. 
For example, this nonclassical degree of freedom \{$b_i$\} plays an important role in the description of phenomena like evaporation, heavy-ion collisions, and fusion (See details in Refs. ~\cite{feldmeier_fermionic_1995,feldmeier_molecular_2000}).

Due to the introduction of  width variables \{$b_i$\} in FMD, one prominent problem is that the spurious center-of-mass effects cannot be treated well especially the time-dependent version of FMD~\cite{feldmeier_molecular_2000}. And the conventional approximate center-of-mass projection technique takes too much numerical effort. Take the  two-nucleon system as a simple example. This problem can serve as how to factorize the following expression in the total center-of-mass motion ($\bm{X}_{\text{cm}}=(\bm{X}_1+\bm{X}_2)/2$) part and the relative motion ($\bm{X}_{\text{rel}}=\bm{X}_2-\bm{X}_1$) part,
\begin{eqnarray}
\exp[-X_1^2/(2b_1^2)-X_2^2/(2b_2^2)]~~~~~~~~~~~~~~~\\
=\exp[-\alpha  \, X_{\text{cm}}^2] \exp[\gamma  \, {\bm{X}_{\text{cm}}}\mycdot \bm{X}_{\text{rel}}]  \exp[-\alpha/4 \, X_{\text{rel}}^2],~
\end{eqnarray}
where $\alpha=(1/b_1^2+1/b_2^2)/2$ and $\gamma=(1/b_1^2-1/b_2^2)/2.$
Clearly, it seems that the only way to remove the spurious center-of-mass cross term $\exp[\gamma \, {\bm{X}_{\text{cm}}}\mycdot \bm{X}_{\text{rel}}] $ is to let the widths $b_1=b_2$, which is just the traditional view for this center-of-mass problem. Kiderlen and Danielewicz~\cite{kiderlen_fragments_1997} ever introduced a width matrix to fix this problem but the antisymmetrization was not treated properly in their method. Indeed,  not only the FMD,  the single-particle basis with the width variable is widely used in nuclear physics but the non-trivial center-of-mass problem is always a dilemma for the many-body trial wave function (e.g., see Refs.~\cite{hagen_solution_2009,feldmeier_molecular_2000}). In shell-model calculations, to remove spurious states of center-of-mass, many specific methods~\cite{Rath_practicalsolutionproblem_1990,Nakada_efficientprojectionalgorithm_1997,Horoi_ExactRemovalCenterofMass_2007} are proposed based on the character of harmonic oscillator basis. For example, the Lawson's method~\cite{GloecknerSpuriouscenterofmassmotion1974} has conventionally used in shell-model calculations, which introduces a shifted center-of-mass Hamiltonian multiplied by a constant parameter. This  method does not require explicit construction of spurious states, nevetheless it cannot be generalized to deal with other many-body wave functions.

In nuclear cluster physics, the same center-of-mass problem arises because of the width variables and actually it is a long-standing problem~\cite{horiuchi_chapter_1977}.  
Firstly, as we know, in the old Resonating Group Method (RGM)~\cite{wildermuth_unified_1977,tang_microscopic_1987}, the trial wave function is expressed in terms of the translationally invariant coordinates and the width variables can be naturally included without the center-of-mass problem. Nevertheless, it is just because of this kind of construction that the practical evaluation is becoming very tedious.  This disadvantage is partly overcame within the Generator Coordinate Method (GCM), in which the many-body wave function can be expressed as the superposition of the Slater determinants. However, when dealing with different width problems, the center-of-mass part cannot be separated in the traditional Brink wave function~\cite{brink_alpha-particle_1966} and other microscopic cluster wave functions~\cite{dufour_molecular_2014}. The double Fourier transformation~\cite{sunkel_generator_1976,horiuchi_chapter_1977,tohsaki-suzuki_new_1978}  from GCM to RGM is a main solution for this width problem. This kind of method is still very complicated and not so practical for calculations. Therefore, in most present GCM calculations, widths of the single-particle basis from different clusters are always assumed to have same values in order to factorize exactly the total center-of-mass motion part, which has been addressed in numerous papers ~\cite{negele_advances_2012}.

Quite recently, originating from the THSR (Tohsaki-Horiuchi-Schuck-R\"{o}pke)  wave function~\cite{tohsaki_alpha_2001,tohsaki_colloquium_2017}, a new container picture~\cite{zhou_nonlocalized_2013,zhou_nonlocalized_2014,funaki_cluster_2015} was proposed for the description of cluster structure in light nuclei. The essential point of the container picture is that a completely new dimension or degree of freedom was clarified, i.e., the size variable \{$B_i$\} or the width variable of the relative Gaussian wave function. It was found that from the typical $n\alpha$ nuclei~\cite{schuck_alpha_2016},  the neutron-rich nuclei~\cite{lyu_investigation_2015,lyu_investigation_2016}, to the hypernuclei~\cite{zhou_container_2014,funaki_container_2014,funaki_cluster_2015}, the size variable \{$B_i$\} can be considered as a true dynamical quantity for describing the correlations of nucleons or clusters,  which goes beyond the traditional inter-cluster distance variable~\cite{funaki_cluster_2015}. 
This introduced size variable \{$B_i$\} is the degree of freedom of the center-of-mass of clusters and this kind of correlation only can be activated by superposing more Slater determinants. Therefore, the width \{$B_i$\} is found to be a new dimension for cluster correlations and it can be considered as a comparatively independent degree of freedom compared with  \{$\bm{Z}_i$\} and  \{$b_i$\} in the many-body wave function. 

In this Letter, I will introduce a new many-body wave function, in which the degrees of freedom \{$b_i$\} and \{$B_i$\} as well as \{$\bm{Z}_i$\} will be employed in a very simple way. Most importantly, the key center-of-mass problem can be solved exactly in the new framework.

I begin with an antisymmetric $A$-nucleon Slater determinant wave function,  
\begin{align}
\label{startwf}
\Phi_0(\bm{r})&=\frac{1}{\sqrt{A!}} {\cal A} [\phi_1(\bm{r}_1) \cdots\phi_{A}(\bm{r}_A)].
\end{align}
Here, the notion $ \bm{r}=\{\bm{r}_1,\cdots,\bm{r}_A \} $ is used. The single-nucleon wave function $\phi_i(\bm{r}_i)$ can be described by a Gaussian wave packet with different widths $b_i$, spins $ \chi_{\sigma_i}$, and isospins  $\chi_{\tau_i} $ variables,
\begin{align}
\phi_i(\bm{r}_i)&=(\frac{1}{\pi b_i^2})^{3/4} e^{- \frac{r_i^2}{2b_i^2}} \otimes \chi_{\tau_i} \otimes \chi_{\sigma_i}.
\end{align}
The nucleons can be distributed into different positions from the origin  by a shift operator $\bm{\widehat{D}}$ defined as, $\bm{\widehat{D}}(\emph{\textbf{R}}) \Phi_0(\emph{\textbf{r}})=\Phi_0(\bm{r}-\bm{R}).$
The obtained $\Phi_0(\bm{r}-\bm{R})$ is nothing new but just the many-body FMD nucleon wave function~\cite{neff_cluster_2004}. 
It is convenient to treat $\Phi_0(\bm{r}-\bm{R})$ as a general cluster wave function for discussing the nucleon wave function and the cluster wave function in a unified way. In this case, the $A$ nucleons can be classified into $n$ groups or clusters and the corresponding generator coordinates are $\bm{R}=\{\bm{R}_1,\cdots,\bm{R}_n \}$~~$(1 \leq  n \leq  A)$.  The $A_i$ nucleons belonging to the $i$th cluster have the same width $b_i$ and they are centered around  $\bm{R}_i$. The nucleon wave function can be included when all the clusters becomes single nucleons in this general cluster wave function. It should be noted that the single-nucleon wave function $\phi_i(\bm{r}_i)$ can also be the higher harmonic oscillator shell-model wave function in Eq.~(\ref{startwf}).  

First, to deal with the center-of-mass problem from the width variables, I construct a many-body integral operator,  
\begin{align}
\bm{\widehat{G}}_n (\beta_0)=\int \! d^3 R_1 \cdots d^3 R_n \text {exp}[{-\sum \limits_{i=1}^n \frac{A_i R_i^2 }{\beta_0^2-2b_i^2} }]\bm{\widehat{D}}(\bm{R}).
\end{align}
With the auxiliary generator coordinate $\bm{R}$,  this $\bm{\widehat{G}}_n (\beta_0)$ operator can perform a simple integral transformation for exactly separating the center-of-mass part from a many-body wave function with or without antisymmetrization.  Take a simple two-cluster example, essentially, the wave function of the center-of-mass motion of two clusters can be factored out as,
\begin{align}
&\bm{\widehat{G}}_2(\beta_0) \, \text{exp}[-A_1 X_1^2/(2b_1^2) -A_2 X_2^2/(2b_2^2) ]\\
&~~~~~~ \propto \text{exp}[ -\frac{A}{\beta_0^2}X_{\text{cm}}^2] \text{exp}[ -\frac{A_1 A_2}{A\beta_0^2} X_{\text{rel}}^2 ],
\end{align}
where the center-of-mass $\bm{X}_{\text{cm}}=(A_1 \bm{X}_1+A_2 \bm{X}_2)/A$ and the relative dynamics coordinate $\bm{X}_{\text{rel}}=\bm{X}_2-\bm{X}_1$.  $\bm{X}_1$ and $\bm{X}_2$ are the center-of-mass coordinates of two clusters, respectively. 

Next, to extend the spirit of the container picture to the description of general nuclear systems, a correlation operator can be created, 
\begin{align}
\label{corrop}
\bm{\widehat{L}}_{n-1} (\bm{\beta})=\int \! d^3 \widetilde{T}_1 \cdots d^3\widetilde{T}_{n-1} \text {exp}[{-\sum \limits_{i=1}^{n-1} \frac{\widetilde{T}_i^2}{\beta_i^2}}] \bm{\widehat{D}}(\bm{T}).
\end{align}
Here, the similar notions $\bm{\beta}=\{\beta_1,\cdots, \beta_{n-1}\}$, $ \bm{T}=\{\bm{T}_1,\cdots,\bm{T}_{n} \} $,
and $\widetilde{\bm T}=\{\widetilde{T}_1,\cdots,\widetilde{T}_{n-1} \} $ are used. It should be noted that $\widetilde{\bm T}$ is the Jacobi coordinate of $ \bm{T}$, 
\begin{align}
\widetilde{\bm T}_k= \bm{T}_{k+1}-  \sum \limits_{i=1}^k A_{i} \bm{T}_i/\sum \limits_{i=1}^{k}A_i.
\end{align}
The total center-of-mass coordinate of the generator coordinate $\bm{T}$ is set to zero. In practical integrations, $ \bm{T}$ can be expressed by $\widetilde{\bm T}$ using the substitution of the variable. The correlation operator in Eq.~(\ref{corrop}) provides us with a very simple way to unfreeze the conventional fixed width variable of the Gaussian relative wave function. 

Finally, complete with building blocks needed and a new many-body wave function is constructed as,
\begin{align}
\Psi_{\text{new}}=& \bm{\widehat{L}}_{n-1} (\bm{\beta}) \bm{\widehat{G}}_n (\beta_0) \bm{\widehat{D}}(\bm{Z}) \Phi_0(\emph{\textbf{r}}) \label{wf1} \\
=&\int \! d^3{\widetilde{T}}_1 \cdots d^3 {\widetilde{T}}_{n-1} \text {exp}[{-\sum \limits_{i=1}^{n-1} \frac{\widetilde{T}_i^2}{\beta_i^2}}] \int \! d^3 {R}_1 \cdots d^3 R_n  \label{wfa} \\
\times&\text {exp}[{-\sum \limits_{i=1}^n (\frac{A_i }{\beta_0^2-2b_i^2})(\bm{R}_i-\bm{Z}_i-\bm{T}_i)^2}]\Phi_0(\bm{r}-\bm{R}) \label{wf2}  \\
=&n_0 \exp[-\frac{A}{\beta_0^2} X_{\text{cm}}^2]~{\cal A} \{ \prod_{i=1}^{n-1}\exp[-\frac{1}{2B_i^2}(\bm{\xi}_i-\bm{S}_i)^2] \prod_{i=1}^{n} \phi_i^\text{int}(b_i)  \}.\label{wf3}
\end{align}
Here $n_0$ is the trivial coefficient factor.  ${\bm X}_{\text{cm}}$ is the total center-of-mass dynamical coordinate. 
The center-of-mass coordinate of generator coordinates $\bm{Z}$ is set to zero. The $\bm{\xi}$ and $\bm{S}$ are the Jacobi coordinates of dynamic coordinate $\bm{X}_i$ and generator coordinate $\bm{Z}_i$, respectively.
The $\phi_i^\text{int}(b_i) $ is the $i$th-cluster intrinsic wave function with the width variable $b_i$ and the spin and isospin variables are also included in it. 
The $B_k$ is the width of the Gaussian relative wave function, 
\begin{align}
B_k^2= \frac{1}{2} [\sum \limits_{i=1}^{k+1}A_i/ (A_{k+1}\sum \limits_{i=1}^k A_{i})] \beta_0^2+\frac{1}{2}\beta_k^2.
\end{align}

\begin{figure}[htbp]
	\begin{center}
		\includegraphics[scale=0.24]{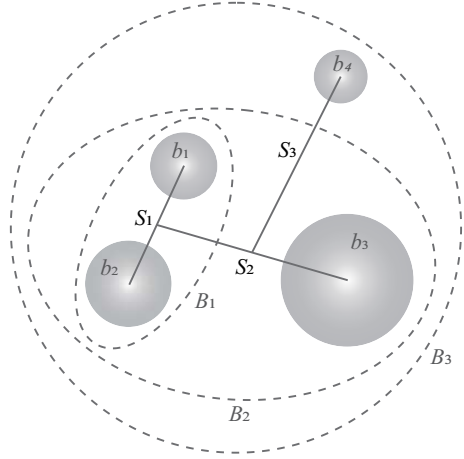}
		\caption{\label{newwf}
			Illustrative diagram for the inter-cluster distance variables \{$\bm{S}_1$, $\bm{S}_2$, $\bm{S}_3$\}, widths of cluster variables  \{${b}_1$, $b_2$, $b_3$, $b_4$\}, and widths of Gaussian relative wave functions  \{${B}_1$, $B_2$, $B_3$\} in the new wave function for a four-cluster system. } 
	\end{center}
\end{figure} 

From the mathematical formulas in Eqs.~(\ref{wf1})$-$(\ref{wf3}) and the corresponding illustrative diagram for a four-cluster system in Fig.~\ref{newwf}, this new wave function is characterized by three important features. 

First and most importantly, despite taking the widths of Gaussian \{$b_i$\} as variables, by using the operator $\bm{\widehat{G}}_n (\beta_0)$  the obtained wave function exactly factorizes to a product of a spurious total center-of-mass wave function and a translation-invariant intrinsic wave function. This fact came as a surprise. There is no doubt that this kind of  exact separation in Eq.~(\ref{wf2}) is the best solution for the center-of-mass problem at present and it breaks the bottleneck for studying many-body effects in nuclear physics taking the width parameters as dynamical variables.

Secondly, by using the correlation operator $\bm{\widehat{L}}_{n-1} (\bm{\beta})$,  the widths of the Gaussian relative wave functions \{$B_i$\} between clusters are treated as variables rather than keeping them at fixed values. After the exact separation of the center-of-mass part, the explicit relative Gaussian wave function $\exp[-(\bm{\xi}_i-\bm{S}_i)^2/(2B_i^2)]$ is clearly manifest, which is the key to explore the correlations of nuclear system. Besides the inter-cluster distance variables $\{\bm{S}_i\}$ (or $\{\bm{Z}_i\}$), other more important width variables \{$B_i$\} are also included, which is one unique feature for this new many-body wave function compared with the FMD and AMD. As mentioned, it is found that in the THSR~\cite{tohsaki_alpha_2001,tohsaki_colloquium_2017} or container picture~\cite{zhou_nonlocalized_2014}, the widths of Gaussian relative wave functions, rather than the traditional inter-cluster distance parameter, are the true dynamical variables for the description of clusters. 

Thirdly, the calculations of matrix elements from this trial wave function are based on the Slater determinant and analytical integration techniques. The introduced variables in Eq.~(\ref{wf3}) have clear physical meanings as we discussed above. Nevertheless it is known that the direct calculations for this kind of antisymmetrical wave function is not realistic. The strategy is that, we can firstly obtain the analytical kernels of the wave function $\Phi_0(\bm{r}-\bm{R})$, which have been studied for many years~\cite{kamimura_new_1974,horiuchi_chapter_1977,tohsaki-suzuki_chapter_1977,beck_calculating_1981}. Then, by meanings of the created  integral formula in Eq.~(\ref{wfa}-\ref{wf2}) multiplying a constructed Gaussian-form factor, high dimensional integrals can be performed analytically, which does not lose any accuracy. This kind of analytical integration method can be found in Refs.~\cite{tohsaki_microscopic_2006,funaki_cluster_2015,myo_tensor-optimized_2015}. Some standard many-body techniques~\cite{ring_nuclear_2004} like the angular-momentum and parity projection and GCM can be directly applied for this wave function.  It should be noted that, for some heavier nuclei like $^{40}$Ca+$\alpha$ system or larger-number clusters system like 5$\alpha$ system, similar with the THSR wave function, this will become a time-consuming computation due to the treatment of huge analytical expressions. To overcome this point, the Monte Carlo technique~\cite{itagaki_three_2010,lyu_investigation_2015} is one possible way for the numerical treatment of the matrix elements for this trail wave function in the future.
Moreover, two simplified versions of this general wave function should be emphasized here. Without considering the correlation operator, $\bm{\widehat{G}}_n (\beta_0) \bm{\widehat{D}}(\bm{Z}) \Phi_0(\emph{\textbf{r}})$ is an interesting FMD-type wave function but the center-of-mass can be removed and the total size of the nucleus is characterized by the parameter $\beta_0$. While without adopting traditional inter-cluster distance variables, the  $\bm{\widehat{L}}_{n-1} (\bm{\beta}) \bm{\widehat{G}}_n (\beta_0) \Phi_0(\emph{\textbf{r}})$ characterized nonlocalized clustering will be another quite interesting THSR-type wave function.

\begin{figure}[htbp]
	\begin{center}
		\includegraphics[scale=0.36]{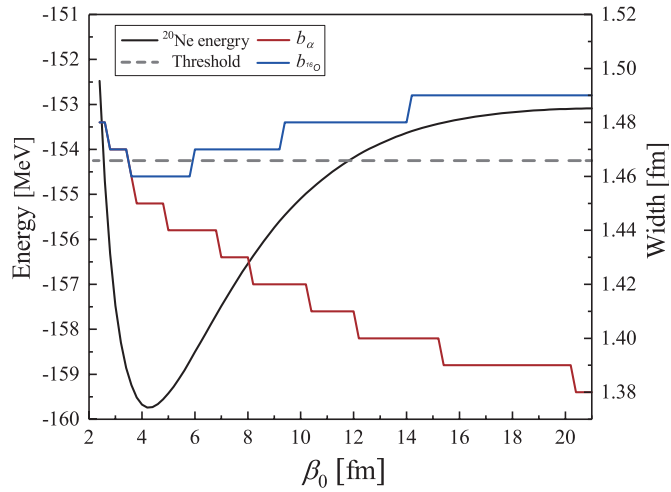}
		\caption{\label{width}Variational energy calculations for \nene\ in the three-parameter space, the width variable of $\alpha$, the width variable of \oo, and the size parameter $\beta_0$, by using Eq.~(\ref{newf}). It should be noted that the widths' lines are not smooth due to the choice of the adopted 0.01-fm meshpoints. }
	\end{center}
\end{figure} 

To test the practicability of the new wave function, as a first step, I will apply this many-body wave function for a cluster system and the nucleus \nene ($\alpha$+\oo) is chosen as a good touchstone.
As we know, \nene\ has a very typical $\alpha$+\oo\ cluster structure and it has been studied for almost half a century~\cite{horiuchi_molecule-like_1968,matsuse_study_1975,zhou_nonlocalized_2014}. However, there still exists no effective and simple solution~\cite{kamimura_new_1974,horiuchi_chapter_1977} for the serious center-of-mass problem even in this two-cluster system with different widths. That is the reason why we also used the same width variable for the $\alpha$ and \oo\ clusters in our recent THSR study for \nene~\cite{zhou_nonlocalized_2013}. In this case,  we indeed assumed the nucleons belonging to $\alpha$ cluster and \oo\ cluster have the same Gaussian wave packets, which is clearly not realistic considering the very different r.m.s radii of the matter distributions of $\alpha$ and \oo\ clusters~\cite{lemere_study_1976}. Now, only by using the $\bm{\widehat{G}}_2(\beta_0)$ operator, the widths of $\alpha$ cluster and \oo\ cluster can be easily treated as variables and some general features are expected to be obtained in the new framework.

Based on the proposed new wave function, the \nene\ wave function can be written directly, 
\begin{eqnarray}  
\label{newf}
\Psi_{\text {Ne}}(\beta_0,b_{\alpha},b_{{}^{16}\text{O}}) = \int \! d^3 R_1 d^3 R_2 \exp[-\frac{4 R_1^2}{ \beta_0^2-2b_{\alpha}^2} -\frac{16 R_2^2}{ \beta_0^2-2b_{{}^{16}\text{O}}^2}] \Phi^\text{B}_{\text {Ne}}(\bm{R}_1,\bm{R}_2)~~~~ \\
\propto \exp[-\frac{20}{\beta_0^2} X_{\text{cm}}^2] ~ {\cal A} \{\exp[-\frac{16}{5\beta_0^2} X_{\text{rel}}^2]~ 
\phi_{\alpha}^{\text{int}}(b_{\alpha}) 
\phi_{{}^{16}\text{O}}^{\text{int}}(b_{{}^{16}\text{O}}) \label{newf2} \}. ~~~~~~~\\
\Phi^\text{B}_{\text {Ne}}(\bm{R}_1,\bm{R}_2)\propto{\cal A} \{\exp[-\frac{2}{b_{\alpha}^2}(\bm{X}_1-\bm{R}_1)^2-\frac{8}{b_{{}^{16}\text{O}}^2} (\bm{X}_2-\bm{R}_2)^2]~ 
\phi_{\alpha}^{\text{int}}(b_{\alpha})
\phi_{{}^{16}\text{O}}^{\text{int}}(b_{{}^{16}\text{O}}) \label{genb}\}.
\end{eqnarray}
Here ${\bm X}_{\text{cm}}=(4 \bm{X}_1+16 \bm{X}_2)/20$ and ${\bm X}_{\text{rel}}=\bm{X}_2-\bm{X}_1$.
$\bm{X}_1$ and $\bm{X}_2$ are the center-of-mass coordinate of $\alpha$ cluster and ${}^{16}\text{O}$ cluster, respectively. 
$b_{\alpha}$ and $b_{{}^{16}\text{O}}$ are the width parameters or the size parameters of $\alpha$ cluster and \oo\ cluster, respectively. 
It should be noted that $\Phi^\text{B}_{\text {Ne}}(\bm{R}_1,\bm{R}_2)$  is a generalized Brink wave function~\cite{brink_alpha-particle_1966,tohsaki-suzuki_chapter_1977} for \nene\ system with different width variables. Again, in the traditional Brink wave function, the widths of different clusters  usually are fixed at the same values, i.e., $b_{\alpha}=b_{{}^{16}\text{O}}$  in Eq.~(\ref{genb}),  to avoid the center-of-mass problem. The same Hamiltonian as in Ref.~\cite{zhou_nonlocalized_2013,zhou_nonlocalized_2014} is used here. For clarity and brevity, we here only focus on the intrinsic states of \nene.
Due to the successful separation of the total center-of-mass part and internal wave function part in Eq.~(\ref{genb}), the matrix elements are all obtained analytically in practical calculations. 

 It is interesting to compare the present THSR-type wave function for \nene\ with other microscopic cluster wave functions. There are three important variables, the width of $\alpha$, the width of \oo, and the size parameter $\beta_0$ in Eq.~(\ref{newf}). The size parameter $\beta_0$ is a new dimensional variable for the description of the correlation of clusters compared with the traditional cluster model like the Brink cluster model.  By studying the inversion doublet bands in \nene, the single THSR wave functions with this key size parameter are almost 100\% equivalent~\cite{zhou_new_2012,zhou_nonlocalized_2013} to the corresponding full solution GCM wave functions of \nene. This gives a strong support that the real dynamical variable for describing the correlations of clusters is the size parameter rather than the conventional  inter-cluster distance parameter.  In RGM, GCM, and other microscopic cluster methods, in principle, the relative wave functions of clusters can be solved exactly but it is not easy to discuss the correlations of clusters because of the complex intrinsic wave functions. In the present wave function in Eq.~(\ref{newf2}), based on a simple and explicit intrinsic wave function, we have a clear container picture~\cite{zhou_nonlocalized_2014} for clarifying the correlations of clusters.
 	
Another practical advantage of this new wave function is the computation performance. Compared with our previous THSR wave function~\cite{zhou_new_2012} for \nene\, i.e., the widths of $\alpha$ and \oo\ take the same value, the change of Eq.~(\ref{newf}) is just one more integral for the generator coordinate $\vect{R}$.  From the view of computation,  the Slater determaint-based calculation is very straightforward, in particularly, compared with the other models and approaches. In RGM, some methods~\cite{kamimura_new_1974,lemere_study_1976} for the \nene\ cluster system have been investigated for a long time and they are still very tedious today especially for the different widths case. One developed  practical method is the double Fourier transformation for the analytical GCM kernels. In Ref.~\cite{tohsaki-suzuki_chapter_1977}, Tohsaki introduced a new eliminating procedure of the spurious center-of-mass motion, which was based on the use of multiple integration, some recurrence formula and the special designed algorithm. In Ref.~\cite{KruglanskiEliminationPauliresonances1992}, M. Kruglanski and D. Baye performed the GCM calculations for $\alpha$+\oo\ cluster system with different widths by using the multiple integrals of the GCM matrix elements. Some similar transformation methods~\cite{sunkel_about_1972,suzuki_resonating_1983,HanckRealgeneratorcoordinatemethod1985} are also developed by other groups. 
Most methods are actually closed related with RGM, in which multiple integration and some other extra treatment for the matrix elements are usually necessary. While in the present method, a simple and explicit intrinsic wave function can be exactly separated by only performing the integral of the real generator coordinates without integrating out the center-of-mass wave function and other complex RGM-based techniques.
 
By making variational calculations for widths $b_{\alpha}$, $b_{{}^{16}\text{O}}$, and $\beta_0$ variables, the energy curve of \nene\ and the variation of width variables can be obtained. Figure~\ref{width} shows the variational energy curve of \nene\ along with the obtained optimum widths $b_{\alpha}$ and $b_{{}^{16}\text{O}}$  using the intrinsic wave function in Eq.~(\ref{newf}).  Generally, it can be seen that with the increase of the size parameter $\beta_0$ indicating the cluster system expands, the gap between $b_{\alpha}$ and $b_{{}^{16}\text{O}}$  is becoming larger and larger. 
Another saying, imagine a situation where the stable double-closed-shell nuclei $\alpha$ and \oo\ are approaching each other to form a stable nuclear state (the ground state of \nene),  in the process Fig.~\ref{width} tells us the  \oo\ is slightly shrinking while the $\alpha$ nucleus is becoming more loosely bound. It has been known that the ground state of \nene\ has a very compact cluster state. In this case, the obtained  minimum energy for the ground state is $-$159.74 MeV. The corresponding wave function is characterized by a rather small value of the size parameter  $\beta_0 $=4.2 fm and slightly different values of widths $b_{\alpha}$=1.45 fm and $b_{{}^{16}\text{O}}$=1.46 fm. 
As a comparison, by using the traditional angular-momentum projected Brink wave function in a variational calculation with the common width parameter and inter-cluster distance parameter $R$, the minimum energy $E$($b_{\alpha}$=$b_{{}^{16}\text{O}}=1.47$ fm, $R$= 3.0 fm)= $-$158.43 MeV is obtained. This is more than 1 MeV higher than the optimum energy obtained by our new intrinsic wave function. 

When $\beta_0$ is very small (\textless 3  fm),  the obtained values of $b_{\alpha}$ and $b_{{}^{16}\text{O}}$  reach exactly the same ones in Fig.~\ref{width}. In this case, it can be proved strictly that~\cite{akaishi_cluster_1987,zhou_nonlocalized_2014} when  $\beta_0 \to$ $\sqrt{2}b_{\alpha}$=$\sqrt{2}$$b_{{}^{16}\text{O}}$, the limit wave function is nothing but the SU(3) shell model wave function. 
In the other limit, when the size parameter $\beta_0$ becomes very large (e.g., more than 20 fm) indicating there are hardly any correlations between clusters,  the obtained optimum values of widths, namely $b_{\alpha}$=1.38 fm and $b_{{}^{16}\text{O}}$=1.49 fm, 
are almost corresponding to the widths of free clusters. Most states of \nene\ including the ground state are between the two limits. Originating from the formation of clustering, the difference of sizes of various nucleons or clusters can be considered as a general feature in light nuclei. 
To give a more accurate description of the $\alpha$+\oo\ cluster structure, the size parameters $b_{\alpha}$, $b_{{}^{16}\text{O}}$, and $\beta_0$ can be treated as three generator coordinates in a general GCM calculations. While this is not easy to be treated in the traditional cluster models, e.g., in RGM the widths of clusters can be different but they usually cannot be superposed as variables. Moreover, because the effect of the breathing mode of clusters can also be described, the effective nucleon-nucleon interactions should be considered carefully. The detailed study for \nene\ in this new framework will be shown in a forthcoming paper.

Generally, with the evolution of light nuclear systems, from shell-model states to gas-like cluster states, the formation of clustering is breaking the smooth conventional mean field in nuclei. Roughly speaking, in a well developed cluster system, the clusters can make nonlocalized motion in a cluster-type mean field characterized by the width parameter of relative Gaussian wave function. Simultaneously, the developed clusters are relatively independent in the nuclear system and nucleons belonging to different clusters can also be restricted by another different mean field inside their clusters, which are characterized as the width variables of clusters. Therefore, not only the size parameter for the relative motion of clusters but also the width of the cluster plays an important role in the description of the light nuclear system. Without considering the width variables, many physical quantities like radius, differential cross section, phase shift actually cannot be well reproduced and this point has been discussed for a long time in the RGM approach~\cite{wildermuth_unified_1977,tang_microscopic_1987,toshitaka_kajino_electromagnetic_1984}.

Finally, a new type of many-body wave function in nuclear physics was proposed in this Letter. The widths and positions of Gaussian packets for the single nucleons and also the widths of Gaussian relative wave functions, all can be treated as variables in this general wave function, in which the long-standing center-of-mass problem has been solved in an exact way. As a nuclear cluster wave function, the gas-like cluster structure, halo nuclei, the distortion effects of clusters in light nuclei are very promising to be investigated in a more realistic way in this new framework.

\ack
I would like to especially thank Prof. Peter Schuck for valuable suggestions and revising carefully the manuscript. The author is obliged to Prof. Hisashi Horiuchi, Prof. Akihiro Tohsaki, and Prof. Masaaki Kimura for their encouragements and suggestions about this work. He also thanks for discussions with Prof. Gerd R\"opke, Prof. Zhongzhou Ren, Prof. Yasuro Funaki, Prof. Chang Xu, and Prof. Taiichi Yamada. This work was supported by JSPS KAKENHI Grant Numbers 17K1426207 (Grant-in-Aid for Young Scientists (B)). Numerical computation in this work was carried out at the Yukawa Institute Computer Facility in Kyoto University.

\bibliographystyle{apsrev4-1}
%
\end{document}